# Nonlinear Localized Exitations And the Dynamics of H-Bonds in DNA

SERGEJ FLACH* AND CHARLES R. WILLIS

Department of Physics, Boston University
590 Commonwealth Avenue, Boston, Massachusetts 02215, USA

* present address: Max-Planck-Institut für Physik Komplexer Systeme, Bayreuther Str. 40 H.16, D-01187 Dresden, Germany.

## 1. INTRODUCTION

The functioning of biomolecules has been a field of very extensive research in modern biology. One of the interesting questions e.g. is the understanding of transcription and copying of DNA molecules. In this contribution we apply recent results we obtained for vibrational energy localization in (arbitrary) nonlinear Hamiltonian lattices to models which might reflect parts of the dynamics of H-bonds in DNA molecules. We show using an abstract level of analysis that there exist generic features of whole classes of models with respect to the existence of (very) localized vibrational excitations. By analyzing the properties of energy (heat) flows in the presence of these localized vibrations we show the localized vibrations to control the heat flow, and thus can define functioning on the level of these physical models. Let us however at the beginning spend a few paragraphs in order to express our views on the advantages and limitations of theoretical physics approaches to the initial problem of functioning in biomolecules.

The formulation of the problem of functioning of biomolecules (let us for clarity constrain ourselves to DNA transcription) in terms of theoretical physics amounts to describing specific nonequilibrium processes of a complex system of many constituents ('particles'). These processes might correspond to certain (nonlinear) dynamic processes within a choosen physical model. Still measurements of the biological system might reveal certain equilibrium properties as well (whatever their connection to the functioning is). A good example is denaturation of DNA, which reminds us of a first order phase transition. Thus we see two approaches in which abstract theory of many-particle systems might help to understand DNA functioning. Both of them start with certain model classes known to be good abstractions of different realizations of condensed matter (here the crucial point is the choice of 'reductionism' of a certain biological constituent to a 'particle', e.g. assuming that a base of a DNA-strand is a rigid body, connected to its neighbours by certain bonds). The first approach leads us to the definition of necessary conditions our models have to have in order to say reveal certain equilibrium properties of DNA. A good example is the work of M. Peyrard and coworkers [1,2], who have studied the denaturation of DNA. After the answer is found (with respect to the necessary conditions) we will have some restrictions on the choice of models and parameters.

The second approach starts again with the choosen model classes (or it might also start from the subclass of models obtained from the analysis using the first approach) and poses the question: 'what are the typical nonlinear excitations my class of models posess, and how can the presence of these excitations in the original DNA molecule contribute to functioning?'. Let us emphasize that the two approaches are not alternative - in fact after obtaining the results from the first approach one has to proceed with the second approach anyway. In other words, the first approach yields certain constraints (with respect to the choice of a model) which have to be obeyed in carrying out the second approach. In order to complete the second approach (assuming that we found some typical or generic nonlinear excitations or processes) a nontrivial feedback with experimental knowledge of functioning of DNA is needed.

In fact there exists a third way, which is solely computer simulation of models closer and closer to the original DNA. Besides the always present computer-assisted limitations, this approach is fundamentally different from the first two described ways, because it does not pretend to understand a physical phenomenon and then to find its trace in the biological system (this will be simply impossible because of the complexity of the biomolecule and its functioning). Still this approach might be as (or even more) successful in answering questions about functioning of say DNA than the first two ways.

In the following we will present our results following the second approach, and of course we will have to stop at the end because of lack of the mentioned feedback. Still we hope that our results and thoughts will help to interprete experimental data and to focus and direct future experimental investigations.

## 2. H-BOND DYNAMICS

### 2.1 Model classes

As indicated above, we will focus on the dynamics of H-bonds between two bases of a basepair in DNA. The simplest approximation of this problem might be to consider the bases as rigid masses, connected with their second partner of a basepair through an interaction potential $V(z)$ which provides bound states, and connected to their neighbour bases along the strand through a second interaction potential $\Phi(z)$. We assume discrete translational invariance of the system (which will not be crucial, but demonstrates the novelty of the results in the best way), i.e. we exclude from our considerations differentiation between different bases and/or disorder. For the moment we consider nearest neighbour interaction along the strand and assign to every rigid mass one degree of freedom $Q_{\pm 1,l}$ where $\pm 1$ indentifies the two different strands and $l$ counts the bases along the strand ($Q$ could be a rotation angle or just a scalar displacement). The interaction potentials are assumed to be given by differences of the involved degrees of freedom $(Q_{+1,l} - Q_{-1,l})$ and $(Q_{\pm 1,l} - Q_{\pm 1,l\pm 1})$. Treating the rigid masses classically we use a transformation of the original base coordinates and separate the Hamiltonian $H$ into a part $H_s$ which describes solely the dynamics of the sum coordinate $Y_l = Q_{+1,l} + Q_{-1,l}$ and a part $H_d$ which describes the dynamics of the difference coordinate $X_l = Q_{+1,l} - Q_{-1,l}$ (for details see [1]):

$$H_s = \sum_l \left[\frac{1}{2}\dot{Y}_l^2 + \Phi(Y_l - Y_{l-1})\right] , \qquad (1)$$

$$H_d = \sum_l \left[\frac{1}{2}\dot{X}_l^2 + V(X_l) + \Phi(X_l - X_{l-1})\right] . \qquad (2)$$

The two interaction potentials are given through Taylor expansions around the stable groundstate of the system:

$$V(z) = \sum_{n=2,3,...} \frac{1}{n!} v_n z^n , \quad \Phi(z) = \sum_{n=2,3,...} \frac{1}{n!} \phi_n z^n . \qquad (3)$$

We have to choose a potential $V(z)$ which should describe the H-bond energy. The physical requirement is that $V(z)$ has a bound state (minimum in $z$) and unbounded states (both because we can break an H-bond by using a finite amount of energy and because of the fact of the denaturation transition). Consequently $V(z)$ will have an energy barrier which separates the bound states from the unbound ones. If the function $V(z)$ is infinitely often differentiable, then it follows that the frequency of oscillation of an imaginary particle in this potential will equal $\sqrt{v_2}$ for infinitely small energies and decrease to zero (continuously) as the energy is increased up to the barrier height.

If we would consider only small amplitude oscillations of our variables $X_l$ around the groundstate $X_l = 0$ of the system, a linearization of the equations of motion $\ddot{X}_l = -\partial H_d/\partial X_l$ yields phonon lattice waves as exact solutions (within the linearization) with the dispersion relation $\omega_q^2 = v_2 + 4\phi_2\sin^2(q/2)$, $0 < q < \pi$. Here $q$ is the phonon wave number. Thus we find an optical phonon band with the property $v_2 \leq \omega_q^2 \leq (v_2 + 4\phi_2)$.

## 2.2 Nonlinear localized excitations

These ingredients are sufficient for us in order to make the following statements. System (2) supports sets of one-parameter families of time-periodic nonlinear localized excitations (NLEs) of the form

$$X_l(t) = X_l(t + 2\pi/\omega_1) , \quad X_{l\to\pm\infty} \to 0 , \quad 0 < \omega_1 < \sqrt{v_2} . \qquad (4)$$

The parameter of each family is the fundamental frequency $\omega_1$. Time-periodic NLEs (4) are *exact* solutions of the equations of motion if the following condition is met: the nonresonance condition $k\omega_1 \neq \omega_q$ has to be fulfilled (here $k$ is an arbitrary integer and $\omega_q$ is an arbitrary phonon frequency). This condition can be certainly fulfilled if $\phi_2 \ll 3/4v_2$ and if $\phi_{n\geq 3}$ are zero or small enough. The parameter region of applicability of our statement might be however even larger. The following reasons can be used in support of our statement. The task of finding time-periodic NLEs can be transformed into the task of solving a set of algebraic equations for the Fourier components $A_{kl}$: $X_l(t) = \sum_k A_{kl}\exp(ik\omega_1 t)$ with $k = 0, \pm 1, \pm 2, ....$ It was shown that in terms of certain separatrix manifolds the problem is generically solvable if the nonresonance condition is met [3]. Furthermore we have shown that the nonresonance condition can be obtained considering certain effective potentials of the original problem [4]. Finally we mention a recent NLE existence proof by MacKay and Aubry for the case of infinitesimal weak coupling $\Phi(z)$, where an analytical continuation from the uncoupled case $\Phi(z) = 0$ was carried out [5].

The time-periodic NLEs decay exponentially fast in space provided $\phi_2 > 0$ [6]. They are exact solutions and thus never 'decay' in time without perturbations. As for perturbations, we have shown that as long as the perturbations are not too strong, time-periodic NLEs can be locally perturbed into quasi-periodic NLEs (meaning that the time-dependence of a given variable is quasiperiodic). These quasiperiodic NLEs are not exact solutions of the equations of motion [6]. Quasiperiodic NLEs will radiate energy and transform either into time-periodic NLEs or follow more complex scenarios [4]. Still quasiperiodic NLEs are meaningful objects, since their lifetimes can easily exceed $10^6$ periods of oscillations [4]. The effects of extended perturbations on time-periodic NLEs will be discussed in detail below.

Time-periodic NLEs can be excited on any lattice site. Because of their robustness to perturbations these excitations can be easily observed in finite temperature simulations of example systems [1,7,8]. One could view NLEs as generalized quasiparticles with finite lifetime (at finite temperatures). On short time scales the quasiparticles behave as being uncoupled. On large time scales the coupling introduces relaxations, which lead to creation and annihilation of NLEs at different lattice sites.

Recently Peyrard and coworkers have proposed a modified model [2,9], which is different from (1),(2) in that the interaction potential $\Phi$ contains terms $(X_l + X_{l-1})$ in (2). Analysis of (2) yields then a remarkable change in the thermodynamic properties of $H_d$ - the denaturation transition becomes exactly a first-order phase transition, whereas without these new terms the denaturation transition is only approximately of first order. As it emerges from the analysis [9], the change in the properties of the transition is sharp, even if the additional new terms are switched on perturbatively. Consequently the properties of single excitations will be only smoothly altered with variation of the perturbation. Thus at least for weak enough perturbations the NLE solutions will survive. Indeed numerical simulations [2] have shown that virtually no changes in the dynamical properties of the system appear as long as one does not consider the transition energy (or temperature) itself. Especially NLE structures have been observed both with or without the additional terms.

Estimations for the parameters of our model can be found in the work of Prohofsky [10] and yield $\phi_2/v_2 \sim 0.03...0.05$. These parameter ranges certainly support the existence of very localized NLEs [4]. In terms of the original physical interpretation of our model this means that DNA dynamics supports localized excitations of H-bonds between bases from a base pair. These excitations will have exponential decay in space (along the strands). By itself this statement does not imply any biological interpretation of our results. Of course we could speculate that an existing NLE might act as a precursor of a local DNA-opening. In fact it is even possible that NLEs exist where one H-bond is periodically opened and closed. Still the functioning of DNA is strongly controlled by certain enzymes. Consequently the most important question is whether it is possible to control the properties of the system in the controlled presence of NLEs. In other words, we think it might be much more reasonable to assume that enzymes interact with DNA strands and create NLEs for special purposes. If this hypothesis has some truth, then we indeed can assign biological functioning to an object which is described within nonlinear physics of many particles.

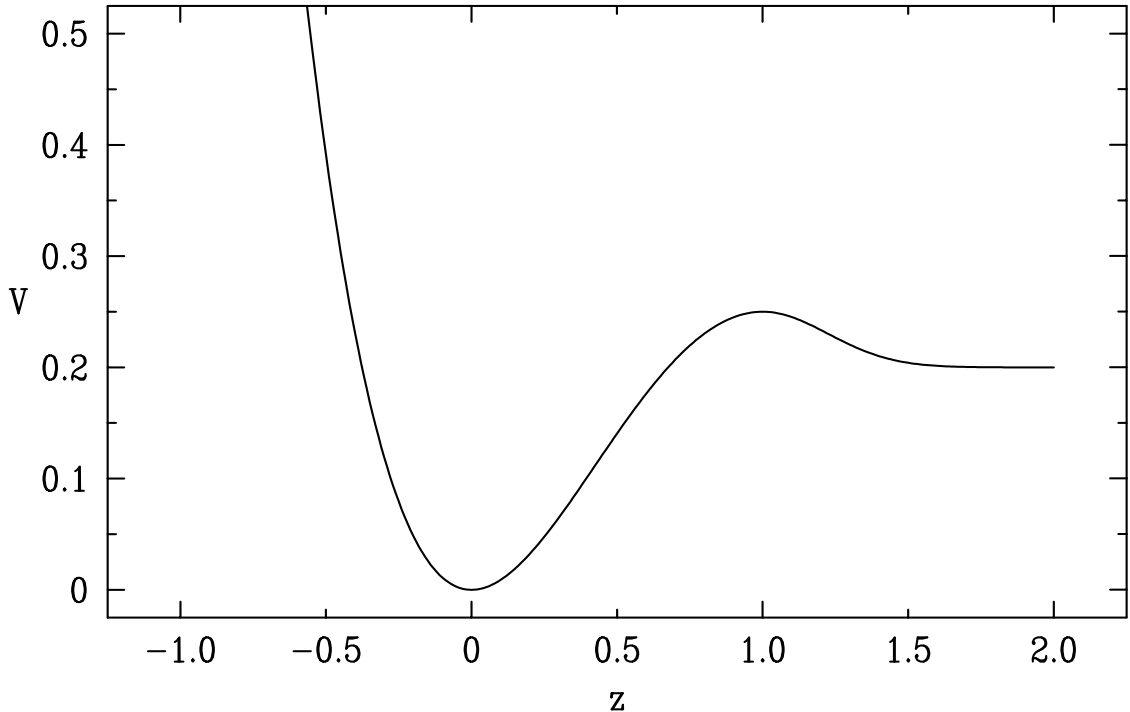

**Fig. 1:** Dependence of the onsite potential $V$ on $z$ (cf. Eq.(5)).

## 3. FUNCTIONING WITH NLEs - PHONON SCATTERING

We can roughly divide the possible functions an NLE fulfills into two classes - biological and physical. The first one is connected with specific chemical and biological processes and is certainly not accessible within our approach. What about the second possibility? Since our toy model is very simple, so are the possible properties of the system. One obvious property is energy or heat transport along the chains (strands). In the following we will show that the presence of an NLE can indeed control the heat flow and thus create a stationary nonequilibrium situation which is not accessible within a usual statistical approach (e.g. of finite temperature simulations).

Let us confine ourselves without any limitation to a simple realization of model (2), such that the details of our results become more concrete. The onsite potential $V(z)$ is choosen as

$$V(z) = \begin{cases} V_{\Phi^4}(z) = \frac{1}{4}((z-1)^2 - 1)^2 & , \quad z \leq 1 \\ V_G(z) = \alpha e^{-\beta(z-1)^2} + \kappa & , \quad z \geq 1 \end{cases} \quad (5)$$

To guarantee smoothness at $z = 1$ up to (including) third derivative we use $0 \leq \kappa \leq 0.25$, $\alpha + \kappa = 0.25$, $2\alpha\beta = 1$. Variations of parameter $\kappa$ are not essential in the following, since we are not interested in the details of say the denaturation transition. The potential $V(z)$ is shown in Fig.1 for $\kappa = 0.2$. The parameter $v_2$ (cf. (3)) is equal to $v_2 = 2$. We choose $\phi_2 = 0.1$, $\phi_{n \neq 2} = 0$ in Eq.(3), so that we are in the parameter range as given in [10]. In that case numerical and analytical studies have revealed the existence of NLE solutions [4,11]. An example for a time-periodic NLE solution is shown in Fig.2 by plotting the Fourier coefficients of every displacement variable $X_l(t)$ versus lattice site. The logarithmic decay in space is clearly observed, in perfect agreement with analytical predictions [6]. These data were obtained using a specific discrete mapping which will be described elsewhere. Quasiperiodic NLE solutions have been also reported [4,11]. Their existence can be visualized with the help of Poincare maps [4]. The properties of NLEs can be studied with the help of reduced problems and effective potentials [4].

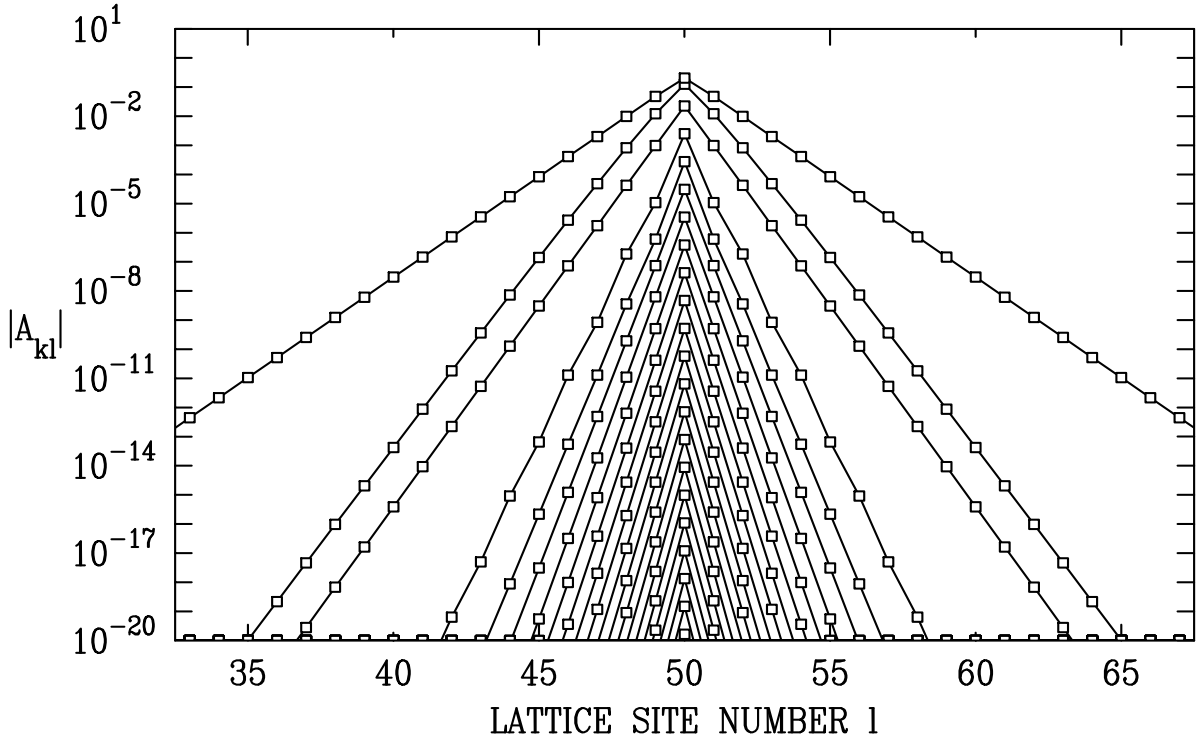

**Fig. 2:** Numerical result of a time-periodic NLE solution with frequency $\omega_1 = 1.3$ for the model parameters as indicated in the text. The absolute values of the Fourier coefficients $A_{kl}$ of the periodic displacements $X_l(t)$ are plotted on a logarithmic scale versus the lattice site $l$. The actual data are represented by open squares. Data for equal Fourier numbers $k$ are connected with solid lines. The Fourier numbers from top to bottom are ($k = 1, 0, 2, 3, 4, 5, ...$).

Let us turn to the question of heat flow control in the presence of a periodic NLE. We create a plane wave of phonons with a given wave number $q$ and small amplitude. We want to measure the transmission coefficient of such a wave by a time-periodic NLE. Since the wave amplitude is assumed to be small, we can account for the scattering problem by adding small perturbations $\delta_l(t)$ to a time-periodic NLE solution $X_l(t)$ and linearizing the equations of motion with respect to the perturbations:

$$\ddot{\delta}_l = -\sum_{l'} \frac{\partial^2 H_{\mathrm{d}}}{\partial X_l \partial X_{l'}}\Big|_{X_l = X_l(t)} \delta_{l'} \quad . \tag{6}$$

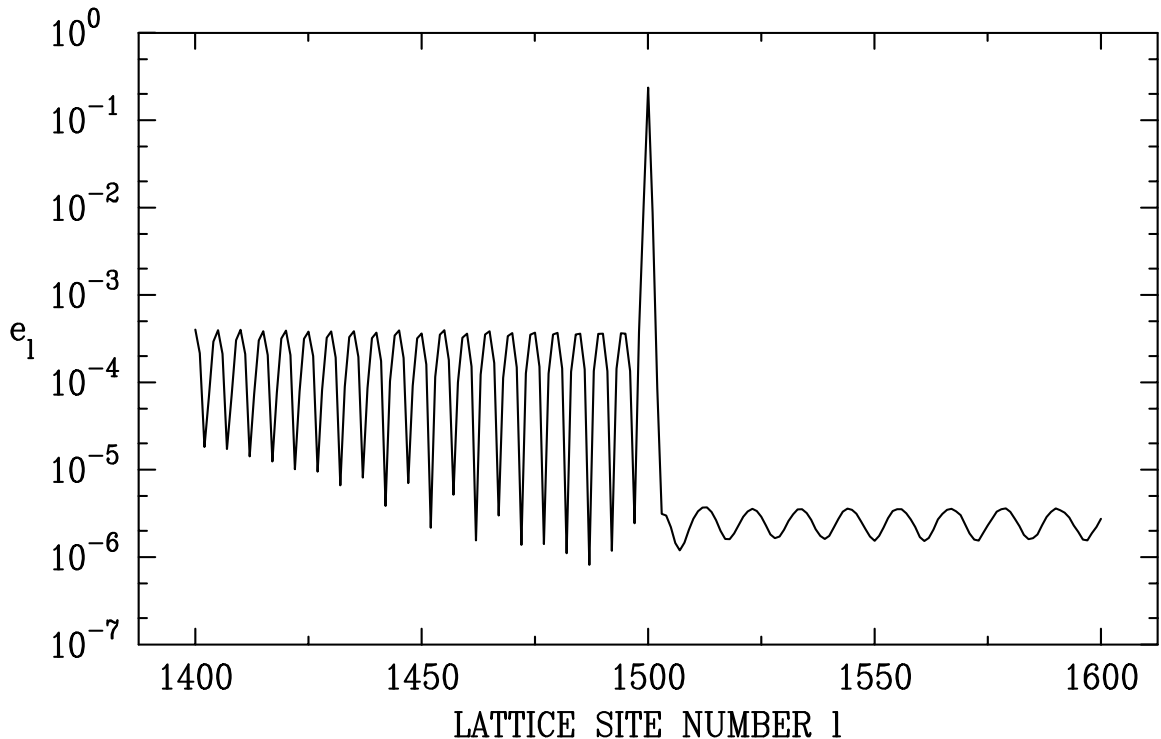

**Fig. 3:** Discrete energy density distribution versus lattice site $l$ of a scattering experiment after a waiting time of $T = 12000$. The infalling phonon wave has energy density $e_l = 10^{-4}$ and wave number $q = 0.2\pi$. The NLE is positioned at $l = 1500$.

Let us first mention, that even if the nonresonance condition for the time-periodic NLE is met, we can not be sure that the NLE is stable in the presence of phonons. As a stability analysis with respect to phonon perturbations shows [4,11], stability of a time-periodic NLE is given if $k\omega_1 \neq 2\omega_q$. In the case of a stable NLE we expect the process of phonon scattering at an NLE to be elastic. The scattering of a phonon wave (6) is then equivalent to the scattering of a single electron in a tight-binding model [12] with a localized array of time-dependent diagonal defects. We do not know at present how to analytically treat this problem of time-dependent multiple scattering. Consequently we will present numerical data of the scattering. We use a time-periodic NLE with energy $E = 0.256$ and frequency $\omega_1 = 1.177$. The phonon waves are created on one side of the NLE and have initial energy density $10^{-4}$ per lattice site. After an initial waiting period (in order to make sure that phonon packets with other wave number than the desired one travel away) we measure the transmitted wave over a time period of $T = 12000$. The system consists of 3000 lattice sites which is sufficiently large that we can exclude reflections from the boundary. The ratio of the transmitted to the incoming energy densities (squared transmission coefficient $|t|^2$) is then measured. In Fig.3 we show the discrete energy density distribution $e_l$ as a function of the lattice site number $l$ after waiting time $T = 12000$:

$$e_l = \frac{1}{2}\dot{X}_l^2 + V(X_l) + \frac{1}{2}(\Phi(X_l - X_{l-1}) + \Phi(X_l - X_{l+1})) \ .$$

Here the NLE is positioned at lattice site $l = 1500$ and the incoming phonon wave ($l < 1500$) with $q = 0.2\pi$ travels to the right and is scattered by the NLE. First we find that the transmitted wave energy density is oscillating in time (as it should be because of the oscillating time-dependence of the diagonal defects in the scattering problem) (see Fig.3, $l > 1500$). Secondly we find that the reflected wave does not loose its coherence with the incoming wave - since we observe standing waves due to the interference between incoming and reflected waves (see Fig.3, $l < 1500$). These standing wave structures are indeed stationary, as analogous pictures for different times clearly show. Note that the ordinate in Fig.3 is plotted on a logarithmic scale. Since the corresponding wave length of the phonon is $\lambda = 10$, one should expect the distance between two neighbouring knots of the standing wave (minima in the energy density) to be $\lambda/2 = 5$. That is precisely observed in Fig.3 ($l < 1500$).

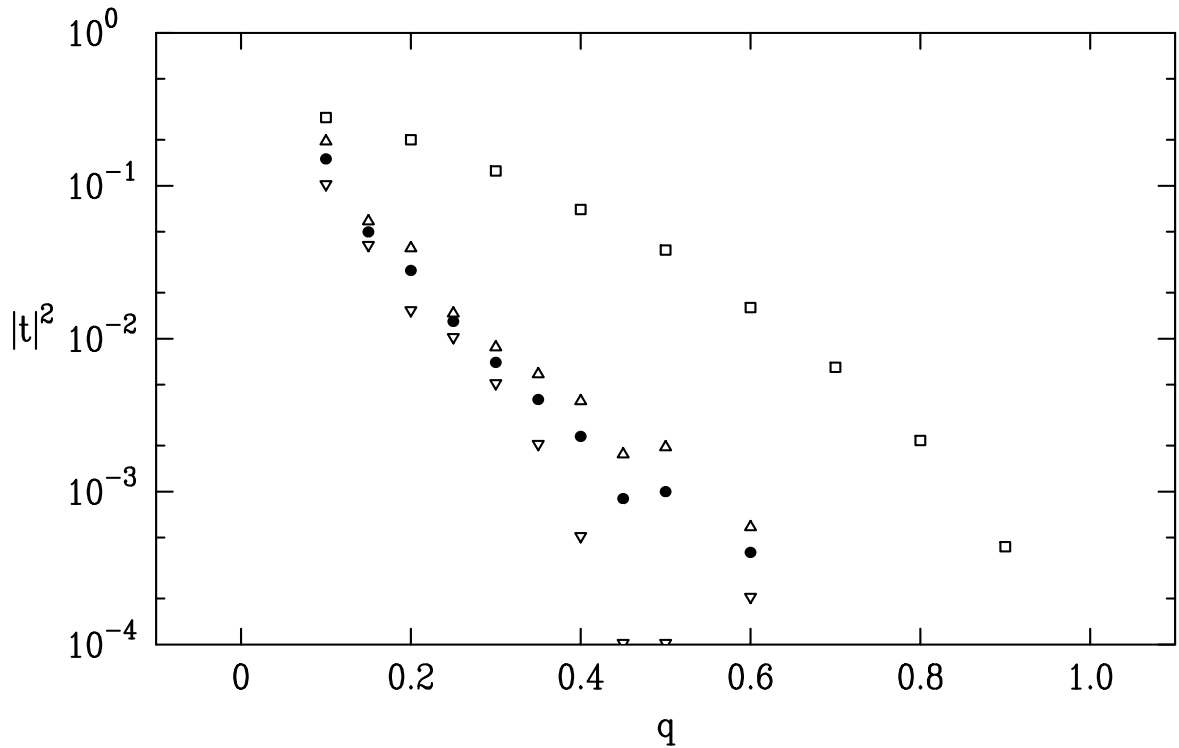

**Fig. 4:** The squared absolute value of the transmittion coefficient $|t|^2$ as a function of wave number $q$ of the infalling phonon wave. The open triangles show the maxima (up) and minima (down) of $|t|^2$, the filled circles show the time-averaged value of $|t|^2$. The open squares correspond to the static linear Ersatzproblem.

But the most important information is the dependence of the squared transmission coefficient on the wave number of the incoming wave as shown in Fig.4. We observe a dramatic decrease of $|t|^2$ over orders of magnitude with increasing wave number $q$. Only for very small wave numbers (large wavelength) do we find appreciable amounts of transmitted wave intensity. This result might appear to be obvious - since large wavelength phonons do not 'feel' the finite size perturbation by the NLE. This is however wrong, as the simple example of *one* static diagonal defect shows - the transmittion coefficient for $q = 0$ is zero as well as for $q = \pi$ (because of the singularities in the phonon density of states) [12].

If we replace the time-dependent defects in equation (6) by their time-averaged val-ues, we obtain a linear Ersatzproblem. The numerical investigation of the phonon scattering for this Ersatzproblem yields the open squares in Fig.4. Allthough the qualitative $q$-dependence seems to be reproduced, the quantitative difference between the time-dependent and the time-independent scattering problems is huge. Consequently we are confronted with a rather subtle result of time-dependent multiple scattering events, which leads,as the numerical result clearly shows, to the fact that phonons with almost any wave number are strongly reflected by a time-periodic NLE. Thus we can prevent the heat flux from penetrating certain areas of our chain by creating two NLEs on the borders of desired region or conversely we can trap energy in a region between two NLEs.. This NLE creation does not require an overall rearrangement of the chain as in the case of a kink but only a local excitation of one or a few degrees of freedom is needed.

If we consider larger interaction ranges than nearest neighbour interaction, we do not expect that the NLE properties will change drastically, as long as we have finite range interaction. As analytical considerations indicate [3,6], there is no basic difference between say next nearest neighbour interaction and nearest neighbour interaction with respect to the existence of NLEs.

Let us return to the problem of DNA functioning. One possible role for NLEs in DNA functioning could be an enzyme which could create NLEs in order to prevent certain areas of DNA from being penetrated by heat flux or to store energy in certain areas. This could be an example of functioning of DNA describable on the level of nonlinear physics of many particles. Of course there can be many more possibilities for functioning using NLEs, however to describe these we would have to leave the abstract level of nonlinear physics and become much more specific in our understanding of the system. It makes no sense to speculate about these possibilities without experimental evidence.

## Acknowledgements

We thank E.W. Prohofsky (Purdue University) for useful discussions and drawing our attention to Ref.[10] and M. Peyrard (E.N.S. Lyon) for many interesting discussions.